\documentclass{raa}

\usepackage{graphicx,times}             
\usepackage{amsmath}
\usepackage{natbib}
\usepackage{url}
\usepackage{amsmath}
\usepackage{amssymb}

\begin{document}

   \title{Astronomical relativistic reference systems with multipolar expansion: the global one\,$^*$
\footnotetext{$*$ Supported by the National Natural Science Foundation of China.}
}
   \volnopage{Vol.0 (200x) No.0, 000--000}      
   \setcounter{page}{1}          

 \author{Yi Xie
   }

   \institute{
             School of Astronomy $\&$ Space Science, Nanjing University, Nanjing 210093, China; {\it yixie@nju.edu.cn}\\
              Key Laboratory of Modern Astronomy and Astrophysics, Nanjing University, Ministry of Education, Nanjing 210093, China
   }

   \date{Received~~ month day; accepted~~~~month day}

\abstract{
With rapid development of techniques for astronomical observations, the precision of measurements has been significantly increasing. Theories of astronomical relativistic reference systems, which are the foundation for processing and interpreting these data now and in the future, may require extensions to satisfy the needs of these trends. Besides building a framework compatible with alternative theories of gravity and the pursuit of higher order post-Newtonian approximation, it will also be necessary to make the first order post-Newtonian multipole moments of celestial bodies be explicitly expressed in the astronomical relativistic reference systems. This will bring some convenience into modeling the observations and experiments and make it easier to distinguish different contributions in measurements. As the first step, the global solar system reference system is multipolarly expanded and the post-Newtonian mass and spin moments are shown explicitly in the metric which describes the coordinates of the system. The full expression of the global metric is given.
}

   \authorrunning{Y. Xie}            
   \titlerunning{Multipolarly expanded global reference system}  

   \maketitle

%
%
\allowdisplaybreaks
\section{Introduction}

Recent years have witnessed the rapid development of techniques for astronomical observations, making the precision of measurements be significantly increasing. One example is the space astrometry mission \textit{Gaia}, which was launched by the European Space Agency (ESA) in 2013 \citep[see][for recent reviews]{Lindegren2008IAUS248.217,Lindegren2010IAUS261.296}. It will obtain accurate astrometric data for $\sim 10^9$ objects from 6th to 20th magnitude. The accuracies for single stars down to 15th magnitude range typically from 8 to 25 micro-arcsecond ($\mu$as). With such a high performance, \textit{Gaia} will be able to detect the relative positional change of a star due to the first order post-Newtonian (1PN) effects from the spherically symmetric parts of gravitational fields of the Sun and some giant planets \citep{Klioner2003AJ125.1580}. In some cases that the observed source is very close to the surfaces of Jupiter and Saturn, the higher order multipole moments of them might cause 1PN light bending up to the level from several tens to hundreds of $\mu$as \citep{Klioner1991SvA35.523,Kopeikin1997JMP38.2587,Klioner2003AJ125.1580}, which are also observable for \textit{Gaia}.

Future space missions may even go further by measuring distances of laser links and angles among these links with unprecedented precision, such as the T\'el\'emetrie InterPlan\'etaire Optique (TIPO) \citep{Samain2002EGSGA27.5808}, the Laser Astrometric Test Of Relativity (LATOR) \citep{Turyshev2004AN325.267}, the Astrodynamical Space Test of Relativity using Optical Devices (ASTROD) \citep{Ni2008IJMPD17.921}, the Search for Anomalous Gravitation using Atomic Sensors (SAGAS) \citep{Wolf2009ExA23.651}, the Phobos Laser Ranging (PLR) \citep{Turyshev2010ExA28.209} and the Beyond Einstein Advanced Coherent Optical Network (BEACON) \citep{Turyshev2009IJMPD18.1025}. Some of them might be able to measure not only 1PN effects caused by the quadrupole moment of the Sun but also effects of the second order post-Newtonian (2PN) light deflection resulting from intrinsic nonlinearity of gravity with high precision.

On the surface of the Earth, time keeping and dissemination equipment are also undergoing great improvements such as optical clocks \citep[e.g.][]{Chou2010PRL104.070802} and optical fiber networks \citep[e.g.][]{Predehl2012Sci336.441}. These technologies will be able to measure the Earth's gravitational potential to new levels of precision by gravitational time dilation at the scale of daily life \citep{Chou2010Sci329.1630} and might bring some subtle effects due to the multipole moments of the Earth into their thresholds in the not-so-far future.

Although, for processing and interpreting these data now and in the future, the International Astronomical Union (IAU) 2000 and subsequent Resolutions\footnote{Resolutions adopted at the IAU General Assemblies: \url{http://www.iau.org/administration/resolutions/general_assemblies/}} on reference systems in the solar system for astrometry, celestial mechanics and metrology in the framework of general relativity (GR) \citep{Soffel2003AJ126.2687} provide a solid foundation, extensions might be required in some specific observations and measurements. To model the light propagation in those observations and experiments accessing 2PN GR effects, some efforts are dedicated to make the IAU Resolutions include all these contributions \citep[e.g.][]{Minazzoli2009PRD79.084027}. Meanwhile, some works are devoted to establish self-consistent astronomical relativistic reference systems compatible with alternative relativistic theories of gravity, such as the scalar-tensor theory \citep{Kopeikin2004PR400.209}, setting up a framework for testing GR. Under these systems, the 2PN theory of light propagation are studied for astronomical observations and the solar system experiments \citep[e.g.][]{Minazzoli2011CQG28.085010,Deng2012PRD86.044007}. Astronomical relativistic reference systems for gravitational sub-systems are also introduced for the advanced theory of lunar motion and for a new generation of lunar laser ranging \citep{Kopeikin2010CMDA108.245,Xie2010APS60.393}.

When higher order post-Newtonian approximation for light propagation is considered, it will also be necessary to make the 1PN multipole moments of celestial bodies be explicitly expressed in the astronomical relativistic reference systems. Because, in some cases like the LATOR mission, the light bending caused by the quadrupole at 1PN order can be comparable with those due to the monopole at 2PN order \citep{Klioner2003AJ125.1580}. It also helps to distinguish effects from the 1PN multipole moments as well as the intrinsic nonlinearity of gravity at 2PN order. However, the IAU Resolutions on astronomical relativistic reference systems are written in the forms of integrals without showing explicit dependence on the mass and spin multipole moments of each local gravitating bodies, which may cause some inconvenience in modeling the observations and experiments and data analysis. To achieve this purpose, it is needed to apply the techniques of multipolar expansion of the gravitational field, which have been intensively studied by many researchers \citep[e.g.][]{Sachs1961RSPSA264.309,Pirani1965LGR.249,Bonnor1966RSPSA289.247,Epstein1975ApJ197.717,Wagoner1979PRD19.2897,Thorne1980RevModPhys52.299,Blanchet1986RSPTA320.379,Blanchet1987RSPSA409.383,Tao1998A&A333.1100}. 

Thus, in this work, I will focus on astronomical relativistic reference systems with multipolar expansion. More specifically, it is to make the 1PN multipole moments be expressed explicitly in the mathematical description of the reference systems within the framework of the scalar-tensor theory. As the first step, the solar system barycentric reference system --- the global one --- will be considered only here. Local reference systems with multipolar expansion will be represented in subsequent works. 

The rest of the paper is organized as follows. Section \ref{sec:RFreview} is devoted to debriefing primary concepts in astronomical relativistic reference systems. In section \ref{sec:multexp}, I present the outline of techniques of multipolar expansion for astronomical relativistic reference systems (a demonstration given in appendix \ref{app:multexpUc}). The full mathematical description of the solar system barycentric reference system with multipolar expansion and its two special cases are shown in appendix \ref{app:exmetric}. Finally, in section \ref{sec:conl}, I summarize the results.

\section{Basics of astronomical relativistic reference systems}

\label{sec:RFreview}

Theories of astronomical relativistic reference systems have been intensively and well studied \citep[e.g.][]{Kopejkin1988CM44.87,Brumberg1989,Brumberg1991Book,DSX1991,DSX1992,DSX1993,Klioner1993A&A279.273,Klioner1993PRD48.1451,Klioner1998A&A334.1123,Kopeikin2004PR400.209,Kopeikin2010CMDA108.245,Xie2010APS60.393}. The following part of this section will give an overview of the primary concepts and necessary mathematical description only \citep[see][for recent reviews and more details]{Kopeikin2011Book,Soffel2013Book}.

A reference system is a mathematical construction which gives ``names'' to spacetime events and a reference frame is a realization of the reference system. A well-defined reference system is the solid and robust foundation for a reference frame which can be materialized by astronomical catalogues and/or dynamical ephemerides of celestial bodies. One leading purpose of classical astrometry in the Newtonian framework is to establish an inertial celestial reference frame. However, this Newtonian absolute space and time is abandoned in GR. In the 4-dimensional curved spacetime, time and space are two parts of a single event. The curvature of spacetime determines motion of matter and the matter, in turn, affects geometry of the spacetime.

An astronomical relativistic reference system is a mathematical description which assigns coordinates (four real numbers) $x^{\mu}$ ($\mu=0,1,2,3$) for an event within it. Among four coordinates, $x^0$ is the time coordinate: $t=c^{-1}x^0$ is the coordinate time where $c$ is the speed of light; and the rest three ones $x^i$ ($i=1,2,3$) are space coordinates. The coordinates $x^{\alpha}=(ct,x^i)$ as a whole are described by the metric tensor $g_{\mu\nu}(x^{\alpha})$ which is a solution of the field equations of Einstein's GR or other alternative relativistic theories of gravity. The metric tensors of reference systems and the coordinate transformations between them hold all of the properties of the reference systems. Although all reference systems are mathematically equivalent, using some specific systems can largely simplify calculations in modeling astronomical and astrophysical processes.

In the solar system, an adequate relativistic description of a gravitational body's motion is not conceivable without a self-consistent theory of astronomical relativistic reference systems. Because the solar system has a hierarchic structure with diversity of various masses of the bodies and the presence of planetary satellite systems which form a set of gravitationally bounded sub-systems. The Sun is the most massive body in the system, but giant planets, like Jupiter and Saturn, can still make it revolve at some distance around the solar system barycenter (SSB). Thus, a global solar system barycentric reference system is required to describe the orbital motion of bodies in the solar system and model the light propagation from distant celestial objects. On the other hand, rotational motion of a body is more natural to describe in the local reference systems associated with each of the bodies. A local reference system of a body is also adequate to describe its figure and satellites' motion. Sometimes, a planet may have natural satellites with non-negligible masses which form a gravitational sub-system. It is convenient to introduce a local reference system associated with the barycenter of the sub-system, which leads to a natural decomposition of orbital motion of the sub-system around SSB and relative motion inside the sub-system.

In 2000, IAU had adopted new resolutions which lay down a self-consistent general relativistic foundation for applications in modern geodesy, fundamental astrometry, celestial mechanics and spacetime navigation in the solar system. These resolutions combine two independent approaches to the theory of relativistic reference systems including the global one and local ones in the solar system developed in a series of publications by Brumberg and Kopeikin (BK formalism) \citep{Kopejkin1988CM44.87,Brumberg1989,Brumberg1991Book} and Damour, Soffel and Xu (DSX formalism) \citep{DSX1991,DSX1992,DSX1993,DSX1994}. 

To make the IAU Resolutions fully compatible with modern ephemerides of the solar system \citep[e.g.][]{Pitjeva2005AstL31.340,Folkner2010IAUS261.155,Fienga2011CMDA111.363} which employ the generalized Einstein-Infeld-Hoffman (EIH) equations \citep{Einstein1938AnMat39.65} with two parameterized post-Newtonian (PPN) parameters $\beta$ and $\gamma$, some efforts \citep{Klioner2000PRD62.024019,Kopeikin2004PR400.209} are contributed. They can go back to the IAU Resolutions when $\beta=1$ and $\gamma=1$. I will follow the approach of \citet{Kopeikin2004PR400.209} in this work.

The metric tensor $g_{\mu\nu}(x^{\alpha})$ under 1PN approximation for any reference system can be formally written as
\begin{eqnarray}
  \label{}
  g_{00} & = & -1+\epsilon^2N+\epsilon^4L+\mathcal{O}(\epsilon^5),\\
  \label{}
  g_{0i} & = & \epsilon^3 L_i+\mathcal{O}(\epsilon^5),\\
  \label{}
  g_{ij} & = & \delta_{ij}+\epsilon^2H_{ij}+\mathcal{O}(\epsilon^4),
\end{eqnarray}
where $\epsilon\equiv1/c$ and $N$, $L$, $L_i$ and $H_{ij}$ are coefficients of the metric. These coefficients can be solved from the field equations of Einstein's GR or other alternative relativistic theories of gravity with certain boundary conditions.

In detail, to solve the metric tensor for the solar system barycentric reference system, it is assumed that the solar system is isolated and there are no masses outside it. The considered number of bodies in the system depends on the required accuracy. Therefore, the spacetime of the solar system is asymptotically flat at infinity with the metric tensor $g_{\mu\nu}$ approaching the Minkowskian metric $\eta_{\mu\nu}=\mathrm{diag}(-1,+1,+1,+1)$. In addition, ``no-incoming-radiation'' conditions are also imposed on the metric tensor to prevent the appearance of non-physical solutions \citep[see][for details]{Kopeikin2004PR400.209}. Its coordinates $x^{\alpha}$ cover the entire spacetime of the solar system and the origin of them coincides with the SSB at any instant of time. The law of conservation of the angular momentum of the solar system can make the spatial axes of the global coordinates non-rotating in space either kinematically or dynamically \citep{Brumberg1989}. A reference system is called kinematically non-rotating if its spatial orientation does not change with respect to the Minkowskian spacetime at infinity as time goes on; a dynamically non-rotating system is defined by the condition that equations of motion of test particles moving with respect to the system do not have any terms that can be interpreted as the Coriolis or centripetal forces. With these assumptions and conditions, the metric tensor $g_{\mu\nu}(x^{\alpha})$ can be obtained in the 1PN approximation within the framework of the scalar-tensor theory \citep{Kopeikin2004PR400.209} and the solutions of its coefficients are given in appendix \ref{app:met} in the form of integrals. Theoretically, this metric can be taken to model observation and experiments; however, its dependence on integrals makes itself inconvenient and non-intuitive in practice. 

Thus, in the next section, these integrals will be multipolarly expanded and expressed in terms of local mass and spin multipole moments of each bodies. This would make the metric tensor easier to deal with and have the physical contribution of multipole moments shown more clearly.

\section{Multipolar expansion of global reference system}

\label{sec:multexp}

To realize the purpose of this work, I need to apply the techniques of relativistic multipolar expansion of the gravitational field, which involves some parameters the so-called mulitpole moments. 

In the Newtonian framework, multipole moments are uniquely defined as coefficients in a Taylor expansion of the gravitational potential in powers of the radial distance from the origin of a reference system to a field point. They can be functions of time in the most general astronomical situations. Multipolar expansion in GR is quite different \citep[see][for a review]{Thorne1980RevModPhys52.299}. Because of the non-linearity of the gravitational interaction, a proper definition of relativistic multipole moments is much more complicated. This issue has been intensively and widely studied \citep[e.g.][]{Sachs1961RSPSA264.309,Pirani1965LGR.249,Bonnor1966RSPSA289.247,Epstein1975ApJ197.717,Wagoner1979PRD19.2897,Thorne1980RevModPhys52.299,Blanchet1986RSPTA320.379,Blanchet1987RSPSA409.383,Tao1998A&A333.1100}. It was shown that, in GR, the multipolar expansion of the gravitational field of an isolated gravitating system is characterized by only two independent sets: mass-type and current-type multipole moments \citep{Thorne1980RevModPhys52.299,Blanchet1986RSPTA320.379,Blanchet1987RSPSA409.383}.

In the scalar-tensor theory of gravity, the multipolar expansion becomes even more complicated due to the scalar field. It introduces an additional set of multipole moments which are caused by the scalar field \citep[see][for details]{Kopeikin2004PR400.209}. In this work, I will follow and apply the techniques of mulitpolar expansion and definitions of multipole moments which have been deeply studied and used in \citet{{Kopeikin2004PR400.209}}.

These required techniques are pretty straightforward. All of the integrals in $g_{\mu\nu}(x^{\alpha})$ [see equations (\ref{app:Uc})--(\ref{app:psiC5}) and (\ref{app:chiC})] for the global reference system can be written in the form as \citep{Kopeikin2004PR400.209}
\begin{equation}
  \label{IntC}
  \mathbf{I}^{(C)}_n\{f\}(t,\bm{x}) = \int_{V_C} f(t,\bm{x}')|\bm{x}-\bm{x}'|^n\mathrm{d}^3x',
\end{equation}
where $n$ is an integer with values of either $-1$ or $1$. It suggests that multipolar expansion of these integrals needs three steps:
\begin{itemize}
	\item Step 1. Taylor expand the integral (\ref{IntC}) using the fact that the characteristic size of the body C is less than the characteristic distance between the field point, $\bm{x}$, and the body C, $\bm{x}_C$, i.e. $|\bm{x}'-\bm{x}_C| < R_C$, where $ \bm{R}_C = \bm{x}-\bm{x}_C$ and $R_C=|\bm{R}_C|$. Here $\bm{x}_C$ represents the position of the center of mass of the body C with respect to the global system and it changes with the global time due to its orbital motion. See Figure \ref{fig:geo} for the geometry of the vectors $\bm{x}$, $\bm{x}'$, $\bm{x}_C$ and $\bm{R}_C$.

	\item Step 2. Convert the global coordinates $\bm{x}'$ of a matter element inside the body C into the local coordinates $\bm{Z}_C'$ with respect to the center of mass of the body C: $\bm{Z}'_C = \bm{x}'-\bm{x}_C+\mathcal{O}(\epsilon^2)$ \citep[see Eq. (11.2.3) in][for details]{Kopeikin2004PR400.209}.  See Figure \ref{fig:geo} for the geometry of the vector $\bm{Z}'_C$.
\begin{figure}[htbp]
  \centering
  \includegraphics[width=12cm]{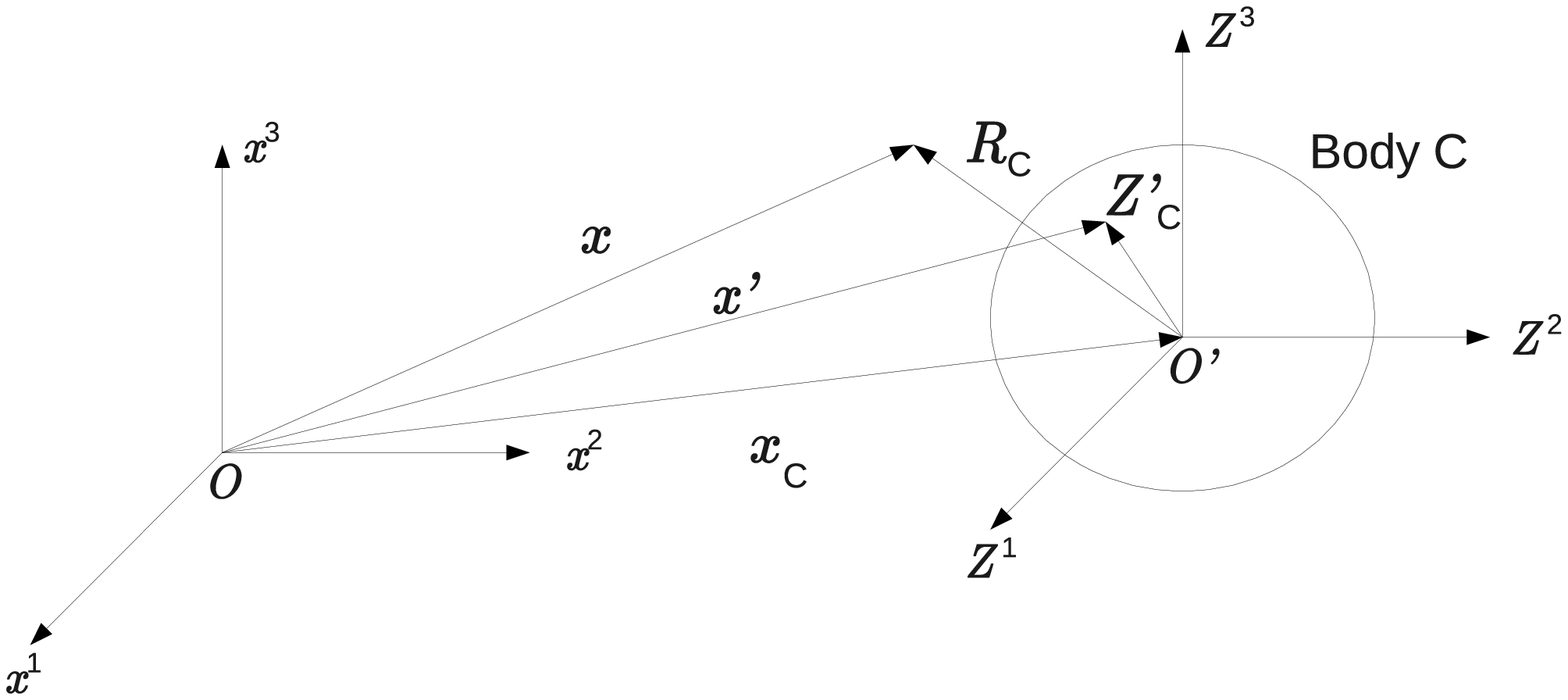}
  \vspace{-0.5cm}
  \caption{The geometry of the vectors $\bm{x}$, $\bm{x}'$, $\bm{x}_C$, $\bm{R}_C$ and $\bm{Z}'_C$.}
  \label{fig:geo}
\end{figure}
	\item Step 3. Collect and rearrange the expansion according to the definitions of mass and spin multipole moments \citep[see Eqs. (6.3.1) and (6.3.8) in][for these definitions]{Kopeikin2004PR400.209}. 
\end{itemize}
To demonstrate this procedure, the multipolar expansion of $U_C(t,\bm{x})$ [see equation (\ref{app:Uc})] is shown in appendix \ref{app:multexpUc} as an example.

After applying it straightforwardly on all of the integrals, the global metric tensor $g_{\mu\nu}$ of the solar system barycentric reference system can be written as
\begin{equation}
  \label{guv}
  g_{\mu\nu} =  \eta_{\mu\nu}+h^{(\mathcal{I})}_{\mu\nu}+h^{(\mathcal{I}^2)}_{\mu\nu}+h^{(\mathcal{S})}_{\mu\nu}+h^{(\mathcal{F})}_{\mu\nu}+h^{(\mathcal{B})}_{\mu\nu} +\mathcal{O}(\epsilon^5),
\end{equation}
where $h^{(\mathcal{I})}_{\mu\nu}$ is contribution from one-body interactions, $h^{(\mathcal{I}^2)}_{\mu\nu}$ originates from two-body interactions, $h^{(\mathcal{S})}_{\mu\nu}$ is due to spins, $h^{(\mathcal{F})}_{\mu\nu}$ contains scaling function $A_C$ and kinematic rotation $F^{km}_C$, and $h^{(\mathcal{B})}_{\mu\nu}$ is caused by bad moments. Their full expressions can be found in appendix \ref{app:exmetric}. $h^{(\mathcal{F})}_{\mu\nu}$ can be eliminated by redefining mass multipole moments and by assuming local reference systems are kinematically non-rotating (see appendix \ref{app:exmetric} for details).  It is worth mentioning that $h^{(\mathcal{B})}_{\mu\nu}$ is gauge-dependent so that it can be eliminated by a coordinate transformation of time component as
\begin{equation}
  \label{}
  t' = t+\epsilon^3\lambda,
\end{equation}
where
\begin{equation}
  \label{}
  \lambda = 2(\gamma+1)\sum_C\sum_{l=0}^{\infty}\frac{(-1)^l(2l+1)}{(2l+3)(l+1)!}G\mathcal{R}_C^{<L>}\bigg(\frac{1}{R_C}\bigg)_{,<L>}.
\end{equation}
Here, $\mathcal{R}_C^{<L>}$ is a so-called ``bad'' moment defined as equation (\ref{BadMR}) \citep{DSX1992} \footnote{Angle brackets surrounding a group of Roman indices denote the symmetric trace-free (STF) part of the corresponding three-dimensional object \citep[see appendix A of][for details]{Blanchet1986RSPTA320.379}. Multi-index notations denotes $L\equiv i_1i_2\cdots i_l$ and comma denotes a partial derivative. Therefore, $Y_{,L}=\partial^{l}Y/\partial x^{i_1}\partial x^{i_2}\cdots\partial x^{i_l}$.}. The components of the new metric $g'_{\mu\nu}$ are
\begin{eqnarray}
  \label{}
  g'_{ij} & = & g_{ij}+\mathcal{O}(\epsilon^4),\\
  g'_{0i} & = & g_{0i}-\epsilon^3\lambda_{,i}+\mathcal{O}(\epsilon^5),\\
  g'_{00} & = & g_{00}-\epsilon^42\lambda_{,t}+\mathcal{O}(\epsilon^5).
\end{eqnarray}
The issue of coordinate transformations and gauges in the relativistic astronomical reference systems is practically important and it has been deeply investigated in several works \citep[e.g.][]{Tao1998A&A333.1100,Tao2000A&A363.335, Tao2006ApSS302.93}

\section{Conclusions}

\label{sec:conl} 

With advances in techniques for astronomical observations and experiments, the theories of astronomical relativistic reference systems might require extensions to satisfy the needs of new high-precision measurements. One direction of them is to make the 1PN multipole moments of celestial bodies be explicitly expressed in the reference systems. Since the effects of both these moments and nonlinearity of gravity are accessible for future space missions, it will bring some convenience for modeling the observations and experiments and make it easier to distinguish different contributions in measurements.

Therefore, as the first step, the global solar system reference system is multipolarly expanded and the 1PN mass and spin moments are made to be shown explicitly in its metric which describes the coordinates of the system. The full expression of the global metric is given (see appendix \ref{app:exmetric} for details of main results). These results might be used in modeling future high-precision time transfer \citep[e.g.][]{Petit1994A&A286.971,Wolf1995A&A304.653,Blanchet2001A&A370.320,Petit2005Metrologia42.138,Nelson2011Metrologia48.171,Deng2012RAA11.703,Deng2013MNRAS431.3236,Deng2013RAA13.1225,Pan2013RAA13.1358,Pan2014RAA14.233}.

\normalem
\begin{acknowledgements}
The work is supported by the National Natural Science Foundation of China (Grant No. 11103010), the Fundamental Research Program of Jiangsu Province of China (Grant No. BK2011553) and the Research Fund for the Doctoral Program of Higher Education of China (Grant No. 20110091120003). I also thank Dr. Sven Zschocke for pointing out a typo in the manuscript.

\end{acknowledgements}

\appendix

\section{metric tensor for global solar system barycentric reference system}

\label{app:met}

The metric tensor $g_{\mu\nu}(x^{\alpha})$ of the global solar system reference system can be solved in the 1PN approximation within the framework of the scalar-tensor theory as \citep{Kopeikin2004PR400.209}
\begin{eqnarray}
  \label{}
  g_{00} & = & -1+\epsilon^2N+\epsilon^4L+\mathcal{O}(\epsilon^5),\\
  \label{}
  g_{0i} & = & \epsilon^3 L_i+\mathcal{O}(\epsilon^5),\\
  \label{}
  g_{ij} & = & \delta_{ij}+\epsilon^2H_{ij}+\mathcal{O}(\epsilon^4),
\end{eqnarray}
where $\epsilon\equiv c^{-1}$ and 
\begin{eqnarray}
  \label{}
  \varphi & = & U(t,\bm{x}),\\	
  \label{}
  N & = & 2\,U(t,\bm{x}),\\
  \label{}
  L & = & 2\Psi(t,\bm{x})-2(\beta-1)\varphi^2(t,\bm{x}) -2U^2(t,\bm{x}) -\frac{\partial^2\chi(t,\bm{x})}{\partial t^2},\\
  \label{}
  L_{i} & = & -2(1+\gamma)\,U_i(t,\bm{x}),\\
  \label{12.7}
  H_{ij} & = & 2\gamma\delta_{ij}U(t,\bm{x}),
\end{eqnarray}
in which $\bm{x}\equiv x^i$ $(i=1,2,3)$ and
\begin{equation}
  \label{}
  \Psi(t,\bm{x})  \equiv  \bigg(\gamma+\frac{1}{2}\bigg)\Psi_1(t,\bm{x})-\frac{1}{6}\Psi_2(t,\bm{x}) +(1+\gamma-2\beta)\Psi_3(t,\bm{x}) +\Psi_4(t,\bm{x})+\gamma\Psi_5(t,\bm{x}),
\end{equation}
Gravitational potentials $U,\,U^i,\,\chi$, and $\Psi_k\,\,$ $(k=1,...,5)$ can be represented as linear combinations of the gravitational potentials of each body of the gravitational system
\begin{equation}
  \label{sumC}
  U=\sum_{C} U_C,\quad U_i=\sum_{C} U^i_C,\quad \Psi_k=\sum_{C}\Psi_{Ck},\quad \chi=\sum_{C}\chi_C,
\end{equation}
where the summation index $C$ numerates the bodies of the system, which gravitational field contributes to the calculations. The gravitational potentials of the body $C$ are defined as integrals taken only over the spatial volume $V_C$ of this body
\begin{eqnarray}
  \label{app:Uc}
  U_{{C}}(t,\bm{x}) & = & G\int_{V_C}\frac{\rho^{\ast}(t,\bm{x}')}{|\bm{x}-\bm{x}'|}\mathrm{d}^3x',\\
  \label{}
  U^i_{{C}}(t,\bm{x}) & = & G\int_{V_C}\frac{\rho^{\ast}(t,\bm{x}')v^i(t,\bm{x}')}{|\bm{x}-\bm{x}'|}\mathrm{d}^3x',\\
  \label{}
  \Psi_{C1}(t,\bm{x}) & = & G\int_{V_C}\frac{\rho^{\ast}(t,\bm{x}')v^2(t,\bm{x}')}{|\bm{x}-\bm{x}'|}\mathrm{d}^3x',\\
  \label{}
  \Psi_{C2}(t,\bm{x}) & = & G\int_{V_C}\frac{\rho^{\ast}(t,\bm{x}')h(t,\bm{x}')}{|\bm{x}-\bm{x}'|}\mathrm{d}^3x',\\
  \label{}
  \Psi_{C3}(t,\bm{x}) & = & G\int_{V_C}\frac{\rho^{\ast}(t,\bm{x}')\varphi(t,\bm{x}')}{|\bm{x}-\bm{x}'|} \mathrm{d}^3x',\\
  \label{}
  \Psi_{C4}(t,\bm{x}) & = & G\int_{V_C}\frac{\rho^{\ast}(t,\bm{x}')\Pi(t,\bm{x}')}{|\bm{x}-\bm{x}'|}\mathrm{d}^3x',\\
  \label{app:psiC5}
  \Psi_{C5}(t,\bm{x}) & = & G\int_{V_C}\frac{\pi^{kk}(t,\bm{x}')}{|\bm{x}-\bm{x}'|}\mathrm{d}^3x',
\end{eqnarray}
where $\rho^{\ast}$ is the invariant density \citep{Fock1959Book}, $\Pi$ is the specific internal energy of matter, $\pi^{\alpha\beta}$ is the anisotropic tensor of stress, $\varphi$ is the perturbation of the scalar field and $h(t,\bm{x})=H_{ii}(t,\bm{x})$. Potential $\chi$ is determined as a particular solution of the inhomogeneous equation
\begin{equation}
  \label{12.18}
  \nabla^2\chi = -2U
\end{equation}
with the right side defined in a whole space and it is
\begin{equation}
  \label{app:chiC}
  \chi_{{C}}(t,\bm{x}) = -G\int_{V_C}\rho^{\ast}(t,\bm{x}')|\bm{x}-\bm{x}'| \mathrm{d}^3x'.
\end{equation}

Mathematically, all of the integrals in the equations (\ref{app:Uc})--(\ref{app:psiC5}) and (\ref{app:chiC}) can be written in the form as \citep{Kopeikin2004PR400.209}
\begin{equation}
  \label{}
  \mathbf{I}^{(C)}_n\{f\}(t,\bm{x}) = \int_{V_C} f(t,\bm{x}')|\bm{x}-\bm{x}'|^n\mathrm{d}^3x',
\end{equation}
where $n$ is an integer with values of either $-1$ or $1$.

\section{Demonstration of Multipolar expansion: the case of $U_C$}

\label{app:multexpUc}

This section of appendixes is devoted to demonstrate the procedure of multipolar expansion for the integrals in equations (\ref{app:Uc})--(\ref{app:psiC5}) and (\ref{app:chiC}). Since it is valid for each of them, we take $U_C$ only as an example and show the details about how to conduct it. For other integrals, what needs is just to repeat it. There are three steps.

\begin{itemize}
  \item Step 1.  Taylor expand the integral (\ref{IntC}) using the fact that the characteristic size of the body C is less than the characteristic distance between the field point, $\bm{x}$, and the body C, $\bm{x}_C$, i.e. $R'_C < R_C$, where $ \bm{R}'_C = \bm{x}'-\bm{x}_C$, $ \bm{R}_C = \bm{x}-\bm{x}_C$ and $R'_C=|\bm{R}'_C|$, $R_C=|\bm{R}_C|$. Here $\bm{x}_C$ represents the position of the center of mass of the body C with respect to the global system.

With the help of
\begin{equation}
  \label{}
   \frac{1}{|\bm{x}-\bm{x}'|}  =   \frac{1}{|\bm{x}-\bm{x}_C-(\bm{x}'-\bm{x}_C)|} = \sum_{l=0}^{\infty}\frac{(-1)^l}{l!}\partial_L\bigg(\frac{1}{R_C}\bigg)R_C'^{<L>},
\end{equation}
in which angle brackets surrounding a group of Roman indices denote the symmetric trace-free (STF) part of the corresponding three-dimensional object \citep[see appendix A of][for details]{Blanchet1986RSPTA320.379} and multi-index notations denote $L\equiv i_1i_2\cdots i_l$ and  $\partial_{L}\equiv \partial_{i_1}\cdots \partial_{i_l}$, the integral in equation (\ref{app:Uc}) can be Taylor expanded as
\begin{equation}
  \label{intrho*}
   \int_{V_C}\frac{\rho^*(t,\bm{x}')}{|\bm{x}-\bm{x}'|}\mathrm{d}^3x' =  \sum_{l=0}^{\infty}\frac{(-1)^l}{l!}\bigg(\frac{1}{R_C}\bigg)_{,L}\int_{V_C} {\rho^*}'R_C'^{<L>}\mathrm{d}^3x',
\end{equation}
where comma denotes a partial derivative.

  \item Step 2. Convert the global coordinates $\bm{x}'$ of a matter element inside the body C into the local coordinates $\bm{Z}_C'$ with respect to the center of mass of the body C.

In the local reference system of the body C, the local coordinates of a field point in the vacuum are $(c\Xi_C,\bm{Z}_C)$ and the coordinates of a matter element inside the body are $(c\Xi_C,\bm{Z}'_C)$, where $\Xi_C$ is its local coordinate time, $\bm{Z}_C$ is the position vector of the field point and $\bm{Z}'_C$ is the position vector pointing at the matter elements. These local coordinates have the relationships with the global coordinates as \citep{Kopeikin2004PR400.209}
\begin{eqnarray}
  \label{}
  \Xi_C & = & t+\epsilon^2\xi^0_C,\\
  Z_C^i & = & R^i_C+\epsilon^2\xi^i_C,\\
  \label{ZC'i}
  Z_C'^i & = & R'^i_C+\epsilon^2[\xi_C'^i+{\mathcal{V}}_C^{\,\prime i}(R_C'^k-R_C^k)v_C^k],
\end{eqnarray}
where $R^i_C = x^i-x^i_C(t)$, $R'^i_C = x'^i-x^i_C(t)$, ${\mathcal{V}}_C^{\,\prime i} = {v'}^i-v_C^i$,
\begin{eqnarray}
  \label{}
  \xi^0_C & = & -(\mathcal{A}_C+v^k_CR_C^k)+\mathcal{O}(\epsilon^2),\\
  \xi^i_C & = & \bigg(\frac{1}{2}v^i_Cv_C^k+D_C^{ik}+F_C^{ik}\bigg)R_C^k +D_C^{ijk}R_C^jR_C^k+\mathcal{O}(\epsilon^2),\\
  \label{xiC'i}
  \xi_C'^i & = & \bigg(\frac{1}{2}v^i_Cv_C^k+D_C^{ik}+F_C^{ik}\bigg)R_C'^k +D_C^{ijk}R_C'^jR_C'^k+\mathcal{O}(\epsilon^2),\\
  F_C^{ij} & = & -\varepsilon^{ijk}\mathcal{F}_C^k,\\
  D_C^{ij} & = & +\delta^{ik}\gamma\bar{U}_C({\bm{x}}_C)-\delta^{ik}A_C,\\
  D_C^{ijk} & = & \frac{1}{2}(a_C^j\delta^{ik}+a_C^k\delta^{ij}-a_C^i\delta^{jk}).
\end{eqnarray}
Therefore, from equations (\ref{ZC'i}) and (\ref{xiC'i}), it has
\begin{eqnarray}
  \label{RC'}
  R_C'^i & = & Z_C'^i -\epsilon^2\bigg[\bigg(\frac{1}{2}v^i_Cv_C^k+D_C^{ik}+F_C^{ik}\bigg)Z_C'^k+D_C^{ijk}Z_C'^jZ_C'^k\bigg]\nonumber\\
  & & +\epsilon^2{\mathcal{V}}_C^{\,\prime i}v_C^k(R_C^k-Z_C'^k)+\mathcal{O}(\epsilon^4),
\end{eqnarray}
which will be a link connecting the global and local coordinates of a matter element.

  \item Step 3. Collect and rearrange the expansion according to the definitions of mass and spin multipole moments.

  According to \citet{Kopeikin2004PR400.209}, the mass multipole moments $\mathcal{I}^{<L>}_{C}$ are defined as
\begin{eqnarray}
  \label{}
  \mathcal{I}^{<L>}_{C}
  & = & \int_{V_{C}}\sigma_{C} (\Xi_C,\bm{Z}_C')Z_C'^{<L>}d^3Z_C' +\frac{\epsilon^2}{2(2l+3)}\bigg[\frac{d^2}{d\Xi_C^2}\int_{V_{C}}\sigma_{C}(\Xi_C,\bm{Z}_C')Z_C'^{<L>}Z_C'^2d^3Z_C'\nonumber\\
  & & -4(1+\gamma)\,\frac{2l+1}{l+1}\frac{d}{d\Xi_C}\int_{V_{C}}\sigma^i_{C}(\Xi_C,\bm{Z}_C')Z_C'^{<iL>}d^3Z_C'\bigg] -\epsilon^2\int_{V_{C}}d^3Z_C'\,\sigma_{C}(\Xi_C,\bm{Z}_C')\nonumber\\
  & & \qquad \times \bigg\{A_{C}+(2\beta-\gamma-1)P_{C} +\sum_{k=1}^{\infty}\frac{1}{k!}\bigg[Q_{C}^{K}+2(\beta-1) P_{C}^{K}\bigg]Z_C'^{K}\bigg\}Z_C'^{<L>},
\end{eqnarray}
in which
\begin{eqnarray}
  \label{}
  \sigma_{C}(\Xi_C,\bm{Z}_C') & = & \rho_C^{\ast}(\Xi_C,\bm{Z}_C')\bigg\{1+\epsilon^2\bigg[\bigg(\gamma+\frac{1}{2}\bigg)\mathcal{V}_C^2(\Xi,\bm{Z}_C') +\Pi_C(\Xi_C,\bm{Z}_C')-(2\beta-1) U_{C}(\Xi_C,\bm{Z}_C')\bigg]\bigg\}\nonumber\\
  & & +\epsilon^2\gamma\pi_C^{kk}(\Xi_C,\bm{Z}_C'),
\end{eqnarray}
and the definitions of the spin moments $S^{<L>}_{C}$ are
\begin{equation}
  \label{}
  S^{<L>}_{C} = \int_{V_{C}}\varepsilon^{pq<i_l} Z_C'^{L-1>p}\sigma^q_{C}(\Xi_C,\bm{Z}_C')d^3Z_C',
\end{equation}
in which $\sigma^i_{C}(\Xi_C,\bm{Z}_C')=\rho_C^{\ast}(\Xi_C,\bm{Z}_C')\mathcal{V}_C^i(\Xi_C,\bm{Z}_C')$.

  With equation (\ref{RC'}) and these two definitions, equation (\ref{intrho*}) can be collected and rearranged further as
\begin{eqnarray}
  \label{}
   & & G\int_C\frac{\rho^*(t,\bm{x}')}{|\bm{x}-\bm{x}'|}d^3x'\nonumber\\
  & = & G\sum_{l=0}^{\infty}\frac{(-1)^l}{l!}\bigg(\frac{1}{R_C}\bigg)_{,L}\bigg\{\mathcal{I}_C^{<L>} +\epsilon^2 \bigg[A_C+(2\beta-\gamma-1)P_C\bigg]\mathcal{I}_C^{<L>}\nonumber\\
  & & -\epsilon^2\int_{V_C} {\rho_C^*}'\bigg[(\gamma+\frac12)\mathcal{V}_C^{\,\prime 2}+\Pi_C'+\gamma\frac{\pi_C^{'kk}}{{\rho_C^*}'} \bigg]Z_C'^{<L>}d^3Z_C' +\epsilon^2(2\beta-1)\int_{V_C} {\rho_C^*}'U_C' Z_C'^{<L>}d^3Z_C'\nonumber\\
  & &-\frac{\epsilon^2}{2(2l+3)}\bigg[\ddot{\mathcal{N}}_C^{<L>}-4(1+\gamma)\,\frac{2l+1}{l+1}\dot{\mathcal{R}}_C^{<L>}\bigg] +\epsilon^2 \sum_{k=1}^{\infty}\frac{1}{k!}Q_C^{K}\int_{C}{\rho_C^*}'Z_C'^{<K>}Z_C'^{<L>}d^3Z_C'\nonumber\\  
   & & +\epsilon^2 2(\beta-1) \sum_{k=1}^{\infty}\frac{1}{k!} P_C^{K}\int_{C}{\rho_C^*}'Z_C'^{<K>}Z_C'^{<L>}d^3Z_C' +\epsilon^2\bigg(-\frac{l}{2}v_C^kv_C^{<i_l}\mathcal{I}_C^{L-1>k} \nonumber\\  
     & & +lF_C^{k<i_l}\mathcal{I}_C^{L-1>k} -lD_C^{k<i_l}\mathcal{I}_C^{L-1>k} -l\mathcal{I}_C^{jk<L-1}D_C^{i_l>jk} -v_C^k\dot{\mathcal{I}}_C^{k<L>}+v_C^kR_C^k\dot{\mathcal{I}}_C^{<L>}\bigg)\bigg\}\nonumber\\
  & & +\epsilon^2G\sum_{l=1}^{\infty}\frac{(-1)^l}{(l+1)!}\bigg(\frac{1}{R_C}\bigg)_{,L}v_C^k\dot{\mathcal{I}}_C^{<kL>} -\epsilon^2G\sum_{l=1}^{\infty}\frac{(-1)^ll}{(l+1)!}\varepsilon_{kpq}v_C^k\bigg(\frac{1}{R_C}\bigg)_{,pL-1}\mathcal{S}_C^{<qL-1>}\nonumber\\
  & & +\epsilon^2G\sum_{l=1}^{\infty}\frac{(-1)^l(2l-1)}{(2l+1)l!}v_C^k\bigg(\frac{1}{R_C}\bigg)_{,kL-1}\mathcal{R}_C^{<L-1>} +\mathcal{O}(\epsilon^4),
\end{eqnarray}
where $\mathcal{N}_C^{<L>}$ and $\mathcal{R}_C^{<L>}$ are called ``bad'' moments defined as \citep{DSX1992}
\begin{equation}
  \label{NL}
  \mathcal{N}_C^{<L>} = \int_{V_C} {\rho_C^*}' Z_C'^2Z_C'^{<L>}d^3Z_C',
\end{equation}
and
\begin{equation}
  \label{BadMR}
  \mathcal{R}_C^{<L>} = \int_{V_C} {\rho_C^*}' \mathcal{V}_C'^kZ_C'^{<kL>}d^3Z_C'.
\end{equation}
This expression is not simplified further because lots of terms will cancel out by terms from other integrals after repeating the same approach and collecting them together.

\end{itemize}

\section{Multipolarly expanded global metric}

\label{app:exmetric}

Full expressions of the global metric $g_{\mu\nu}$ with multipolar expansion can be written as
\begin{eqnarray}
  \label{}
  g_{00} & = & -1+h^{(\mathcal{I})}_{00}+h^{(\mathcal{I}^2)}_{00}+h^{(\mathcal{S})}_{00}+h^{(\mathcal{F})}_{00}+h^{(\mathcal{B})}_{00} +\mathcal{O}(\epsilon^5),\\
  g_{0i} & = & h^{(\mathcal{I})}_{0i}+h^{(\mathcal{S})}_{0i}+h^{(\mathcal{B})}_{0i} +\mathcal{O}(\epsilon^5),\\
  g_{ij} & = & \delta_{ij}+ h^{(\mathcal{I})}_{ij}+\mathcal{O}(\epsilon^4),
\end{eqnarray}
where
\begin{eqnarray}
  \label{}
  h^{(\mathcal{I})}_{00}
& = & \epsilon^22\sum_{C}\sum_{l=0}^{\infty}\frac{(2l-1)!!}{l!}\frac{G\mathcal{I}_C^{<L>}}{R_C^{2l+1}}R_C^{<L>}
   +\epsilon^4\sum_C\sum_{l=0}^{\infty}\frac{[2(2\gamma+1)l+6\gamma+5](2l-1)!!}{(2l+3)l!} \frac{G\mathcal{I}_C^{<L>}}{R_C^{2l+1}}v_C^2R_C^{<L>}\nonumber\\
  & & -\epsilon^4\sum_C\sum_{l=0}^{\infty}\frac{(2l+1)!!}{l!}\frac{G\mathcal{I}_C^{<L>}}{R_C^{2l+3}}v_C^{k}v_C^{m}R_C^{<kmL>} -\epsilon^4\sum_{C}\sum_{l=0}^{\infty}\frac{(2l+1)(2l+1)!!}{(2l+5)l!}\frac{G\mathcal{I}_C^{<kL>}}{R_C^{2l+3}}v_C^{k}v_C^{m}R_C^{<mL>}\nonumber\\
  & & +\epsilon^44(\gamma+1) \sum_{C}\sum_{l=0}^{\infty}\frac{(2l-1)!!}{(l+1)!}\frac{G\dot{\mathcal{I}}_C^{<kL>}}{R_C^{2l+1}}v_C^kR_C^{<L>} -\epsilon^4\sum_C\sum_{l=0}^{\infty}\frac{(2l-3)!!}{l!}\frac{G\ddot{\mathcal{I}}_C^{<L>}}{R_C^{2l-1}}R_C^{<L>},\\
   \label{}
   h^{(\mathcal{I}^2)}_{00}
  & = & -\epsilon^4 2\beta \sum_{C}\sum_{D}\sum_{l,k=0}^{\infty}\frac{(2l-1)!!(2k-1)!!}{l!k!} \frac{G^2\mathcal{I}_C^{<L>}\mathcal{I}_D^{<K>}}{R_C^{2l+1}R_D^{2k+1}}R_C^{<L>} R_D^{<K>}\nonumber\\
  & & -\epsilon^4 2\gamma \sum_{C}\sum_{D\neq C}\sum_{l,k=0}^{\infty}\frac{(l+1)(2l-1)!!(2k-1)!!}{l!k!} \frac{G^2\mathcal{I}_C^{<L>}\mathcal{I}_D^{<K>}}{R_C^{2l+1}R_{CD}^{2k+1}}R_C^{<L>}R_{CD}^{<K>}\nonumber\\
  & & +\epsilon^4\sum_C\sum_{D\neq C}\sum_{l,k,p=0}^{\infty}\frac{(-1)^{p}(2l-1)!!(2k+2p+1)!!}{l!k!p!\mathcal{M}_C}\frac{G^2\mathcal{I}_C^{<L>}\mathcal{I}_C^{<P>}\mathcal{I}_D^{<K>}}{R_C^{2l+1}R_{CD}^{2k+2p+3}}R_C^{<mL>}R_{CD}^{<mKP>}\nonumber\\
  & & +\epsilon^4 2\sum_{C}\sum_{D\neq C}\sum_{l,k,p=0}^{\infty}\frac{(-1)^p(l+2)(2l+1)(2l-1)!!(2k+2p+1)!!}{(2l+3)l!k!p!\mathcal{M}_C} \frac{G^2\mathcal{I}_C^{<mL>}\mathcal{I}_C^{<P>}\mathcal{I}_D^{<K>}}{R_C^{2l+1}R_{CD}^{2k+2p+3}} R_C^{<L>}R_{CD}^{<mKP>},\nonumber\\\\
  \label{}
  h^{(\mathcal{S})}_{00} & = & -\epsilon^44(\gamma+1)\sum_{C}\sum_{l=0}^{\infty}\frac{(2l+1)!!}{(l+2)l!}\frac{G\mathcal{S}_C^{<qL>}}{R_C^{2l+3}} \varepsilon_{kpq}v_C^kR_C^{<pL>},\\
  \label{}
  h^{(\mathcal{F})}_{00}
& = & \phantom{+}\epsilon^4 2\sum_{C}\sum_{l=0}^{\infty}\frac{(l+1)(2l-1)!!}{l!}\frac{G\mathcal{I}_C^{<L>}}{R_C^{2l+1}}A_CR_C^{<L>} +\epsilon^42 \sum_{C}\sum_{l=0}^{\infty}\frac{(2l+1)!!}{l!}\frac{G\mathcal{I}_C^{<kL>}}{R_C^{2l+3}}F_C^{km}R_C^{<mL>},\nonumber\\\\
  \label{}
   h^{(\mathcal{B})}_{00}
  & = & \epsilon^44(\gamma+1)\sum_{C}\sum_{l=0}^{\infty}\frac{(2l+1)(2l+1)!!}{(2l+3)(l+1)!}\frac{G\mathcal{R}_C^{<L>}}{R_C^{2l+3}}v_C^kR_C^{<kL>} +\epsilon^44(\gamma+1)\sum_{C}\sum_{l=0}^{\infty}\frac{(2l+1)!!}{(2l+3)(l+1)!}\frac{G\dot{\mathcal{R}}_C^{<L>}}{R_C^{2l+1}}R_C^{<L>},\nonumber\\\\
  \label{}
  h^{(\mathcal{I})}_{0i}
  & = & -\epsilon^3 2(\gamma+1) \sum_{C}\sum_{l=0}^{\infty}\frac{(2l-1)!!}{l!}\frac{G\mathcal{I}_C^{<L>}}{R_C^{2l+1}}R_C^{<L>}v_C^i -\epsilon^3 2(\gamma+1)\sum_{C}\sum_{l=0}^{\infty}\frac{(2l-1)!!}{(l+1)!}\frac{G\dot{\mathcal{I}}_C^{<iL>}}{R_C^{2l+1}}R_C^{<L>},\\
  \label{}
  h^{(\mathcal{S})}_{0i}
   & = & \epsilon^3 2(\gamma+1)\sum_{C}\sum_{l=0}^{\infty}\frac{(2l+1)!!}{(l+2)l!}\frac{G\mathcal{S}_C^{<qL>}}{R_C^{2l+3}}\varepsilon_{ipq}R_C^{<pL>},\\ 
  \label{}
  h^{(\mathcal{B})}_{0i}
  & = &  -\epsilon^3 2(\gamma+1)\sum_{C}\sum_{l=0}^{\infty}\frac{(2l+1)(2l+1)!!}{(2l+3)(l+1)!}\frac{G\mathcal{R}_C^{<L>}}{R_C^{2l+3}}R_C^{<iL>}, \\ 
  \label{}
   h^{(\mathcal{I})}_{ij}
  & = & \epsilon^22\gamma\delta_{ij}\sum_{C}\sum_{l=0}^{\infty}\frac{(2l-1)!!}{l!}\frac{G\mathcal{I}_C^{<L>}}{R_C^{2l+1}}R_C^{<L>},  
\end{eqnarray}
where $l!!$ means the double factorial of $l$, $\delta_{ij}$ is Kronecker symbol, $\varepsilon_{ijk}$ is Levi-Civita symbol and dot means derivative with respect to time. Since these dots appears at 1PN order only, their difference between the derivative with respect to the global time and that against local times is at the 2PN order. If new mass multipole moments are defined as
\begin{equation}
  \label{}
  \mathcal{I}_{C}^{<L>}\bigg|_{\mathrm{new}} = [1+\epsilon^2 (l+1) A_C]\mathcal{I}_C^{<L>},
\end{equation}
then the first term of $h^{(\mathcal{F})}_{00}$ can be absorbed by the first term of $h^{(\mathcal{I})}_{00}$ and the rest parts of $g_{\mu\nu}$ remain formally unchanged. If the local reference system associated with the body C is kinematically non-rotating, i.e. $F_C^{km}=0$, the second term of $h^{(\mathcal{F})}_{00}$ vanishes.

\subsection{Special cases}

Two special cases can be obtained:
\begin{enumerate}
	\item A single body with arbitrary mass and spin multipole moments. Its metric tensor $g_{\mu\nu}^{(1)}$ reads as 
\begin{eqnarray}
  \label{}
  g_{00}^{(1)}
& = & -1+\epsilon^22\sum_{l=0}^{\infty}\frac{(2l-1)!!}{l!}\frac{G\mathcal{I}^{<L>}}{R^{2l+1}}R^{<L>} -\epsilon^4\sum_{l=0}^{\infty}\frac{(2l-3)!!}{l!}\frac{G\ddot{\mathcal{I}}^{<L>}}{R^{2l-1}}R^{<L>}\nonumber\\
  & & -\epsilon^4 2\beta \sum_{l,k=0}^{\infty}\frac{(2l-1)!!(2k-1)!!}{l!k!}\frac{G^2\mathcal{I}^{<L>}\mathcal{I}^{<K>}}{R^{2l+1}R^{2k+1}} R^{<L>} R^{<K>}\nonumber\\
  & & +\epsilon^44(\gamma+1)\sum_{l=0}^{\infty}\frac{(2l+1)!!}{(2l+3)(l+1)!}\frac{G\dot{\mathcal{R}}^{<L>}}{R^{2l+1}}R^{<L>} +\mathcal{O}(\epsilon^5), \\ 
  \label{}
  g_{0i}^{(1)} 
  & = & -\epsilon^3 2(\gamma+1)\sum_{l=0}^{\infty}\frac{(2l-1)!!}{(l+1)!}\frac{G\dot{\mathcal{I}}^{<iL>}}{R^{2l+1}}R^{<L>} +\epsilon^3 2(\gamma+1)\sum_{l=0}^{\infty}\frac{(2l+1)!!}{(l+2)l!}\frac{G\mathcal{S}^{<qL>}}{R^{2l+3}}\varepsilon_{ipq}R^{<pL>}\nonumber\\
  & & -\epsilon^3 2(\gamma+1)\sum_{l=0}^{\infty}\frac{(2l+1)(2l+1)!!}{(2l+3)(l+1)!}\frac{G\mathcal{R}^{<L>}}{R^{2l+3}}R^{<iL>} +\mathcal{O}(\epsilon^5),\\  
  \label{}
  g_{ij}^{(1)}
  & = &  \delta_{ij}+\epsilon^22\gamma\delta_{ij}\sum_{l=0}^{\infty}\frac{(2l-1)!!}{l!}\frac{G\mathcal{I}^{<L>}}{R^{2l+1}}R^{<L>} +\mathcal{O}(\epsilon^4).  
\end{eqnarray}
When $\gamma=\beta=1$, it can return to the previous results \citep[e.g.][]{Blanchet1986RSPTA320.379,Blanchet1987RSPSA409.383}. This case might be a good approximation for modeling some measurements of LATOR mission after cutting the summations of $l$ to certain values, since LATOR will be able to measure the 1PN effects caused by the quadrupole moment of the Sun \citep{Turyshev2004AN325.267}.

	\item A system consisting of $N$ spinning point masses. Its metric tensor $g_{\mu\nu}^{(N)}$ has the form as
\begin{eqnarray}
  \label{}
  g_{00}^{(N)} 
  & = & -1 +\epsilon^22\sum_{C}\frac{G\mathcal{M}_C}{R_C}
   +\epsilon^42(\gamma+1)\sum_C\frac{G\mathcal{M}_C}{R_C}v_C^2 -\epsilon^4\sum_C\frac{G\mathcal{M}_C}{R_C^{3}}(v_C^{k}R_C^{k})^2\nonumber\\
  & & -\epsilon^4 2\beta \sum_{C}\sum_{D}\frac{G^2\mathcal{M}_C\mathcal{M}_D}{R_CR_D} -\epsilon^4 2\gamma \sum_{C}\sum_{D\neq C}\frac{G^2\mathcal{M}_C\mathcal{M}_D}{R_CR_{CD}}\nonumber\\
  & & +\epsilon^4\sum_C\sum_{D\neq C}\frac{G^2\mathcal{M}_C\mathcal{M}_D}{R_CR_{CD}}R_C^{m}R_{CD}^{m}
   -\epsilon^42(\gamma+1)\sum_{C}\frac{G\mathcal{S}_C^{q}}{R_C^{3}}\varepsilon_{kpq}v_C^kR_C^{p}\nonumber\\
  & & +\epsilon^4 2\sum_{C}\frac{G\mathcal{M}_C}{R_C}A_C +\mathcal{O}(\epsilon^5), \\ 
  \label{}
  g_{0i}^{(N)} 
  & = & -\epsilon^3 2(\gamma+1) \sum_{C}\frac{G\mathcal{M}_C}{R_C}v_C^i +\epsilon^3 (\gamma+1)\sum_{C}\frac{G\mathcal{S}_C^{q}}{R_C^{3}}\varepsilon_{ipq}R_C^{p} +\mathcal{O}(\epsilon^5),  \\
  \label{}
  g_{ij}^{(N)} 
   & = & \delta_{ij}+\epsilon^22\gamma\delta_{ij}\sum_{C}\frac{G\mathcal{M}_C}{R_C}
   +\mathcal{O}(\epsilon^4).  
\end{eqnarray}
If a sub-case is considered that $\gamma=\beta=1$, $g_{\mu\nu}^{(N)}$ identically matches the global metric shown in previous works \citep{Will1993TEGP,Soffel2003AJ126.2687}.
\end{enumerate}

\bibliographystyle{raa.bst}
\bibliography{Gravity20140707.bib}

\end{document}